\documentclass[conference]{IEEEtran}
\IEEEoverridecommandlockouts
\usepackage{cite}
\usepackage{amsmath,amssymb,amsfonts}
\usepackage{algorithmic}
\usepackage{graphicx}
\usepackage{textcomp}
\usepackage{xcolor}
\usepackage{url}
\usepackage{tcolorbox}
\def\BibTeX{{\rm B\kern-.05em{\sc i\kern-.025em b}\kern-.08em
    T\kern-.1667em\lower.7ex\hbox{E}\kern-.125emX}}

\begin{document}

\title{Cybersecurity Defenses: Exploration of CVE Types through Attack Descriptions
}

\author{\IEEEauthorblockN{Refat Othman}
\IEEEauthorblockA{\textit{Faculty of Engineering} \\
\textit{Free University of Bozen-Bolzano}\\
Bolzano, Italy \\
ramneh@unibz.it}
\and
\IEEEauthorblockN{Bruno Rossi}
\IEEEauthorblockA{\textit{Faculty of Informatics} \\
\textit{Masaryk University}\\
Brno, Czech Republic \\
brossi@mail.muni.cz}
\and

\IEEEauthorblockN{Barbara Russo}
\IEEEauthorblockA{\textit{Faculty of Engineering} \\
\textit{Free University of Bozen-Bolzano}\\
Bolzano, Italy \\
brusso@unibz.it}

}

\maketitle

\begin{abstract}
% In the realm of software security, vulnerabilities in software programs can remain undetected for long even after attackers exploit them. 
% Thus, associating attacks to vulnerabilities would help security experts to promptly identify them and respond to the incident. 
% This paper introduces  VULDAT, a classification tool that utilizes MPNET, a sentence transformer, to automatically identify system vulnerabilities from the description of attack techniques. We applied VULDAT to 100 attack techniques collected from the MITRE ATT\&CK repository and 685 issues of the MITRE CVE repository. Then, we compare the performance and accuracy of VULDAT against the other eight SOTA classifiers based on sentence transformers.  
% Our findings indicate that VULDAT achieves the best F1 score of 0.85, Precision of 0.86, and Recall of 0.83. Furthermore, we  found that the vulnerabilities  in the CVE repository and associated to an attack include on average 56\% of the vulnerabilities identified by VULDAT and vice-versa, the vulnerabilities detected by VULDAT for an attack include on average 61\% of the vulnerabilities of the CVE repository. 
% % Upon manual validation, it was observed that VULDAT has the potential to enhance the MITRE knowledge base and identify new vulnerabilities from attack techniques, marking a promising advancement in the field.
% Thus, VULDAT is able to associate attack techniques with vulnerabilities, enhance the detection and response to software security incidents, and thereby contribute to more secure software systems.
Vulnerabilities in software security can remain undiscovered even after being exploited. Linking attacks to vulnerabilities helps experts identify and respond promptly to the incident.
This paper introduces VULDAT, a classification tool using a sentence transformer MPNET to identify system vulnerabilities from attack descriptions. Our model was applied to 100 attack techniques from the ATT\&CK repository and 685 issues from the CVE repository. Then, we compare the performance of VULDAT against the other eight state-of-the-art classifiers based on sentence transformers.  
Our findings indicate that our model achieves the best performance with F1 score of 0.85, Precision of 0.86, and Recall of 0.83. Furthermore, we found 56\% of CVE reports vulnerabilities associated with an attack were identified by VULDAT, and 61\% of identified vulnerabilities were in the CVE repository.
% VULDAT enhances detection and response to software security incidents, contributing to more secure systems.

\end{abstract}
\enlargethispage{\baselineskip}

%\vspace{-0.2cm}
\section{Introduction}
\enlargethispage{\baselineskip}
Cyberattacks are expected to cost organizations \$3 trillion in 2015, \$6 trillion in 2021, and more than \$10.5 trillion annually by 2025~\cite{Cybercrime}. Moreover, an attacker targets the organization's system over a thousand times a week on average~\cite{CResearch}. However, Cyber Threat Intelligence (CTI) can be utilized by cybersecurity practitioners~\cite{rahman2023attackers}~\cite{othman2023vuldat}, which refers to the information that practitioners can use to protect themselves proactively against cyberattacks.
% In this context, there is a necessity for better cyber information sources and a standardized cybersecurity knowledge database. It is necessary to use cybersecurity knowledge databases and cyber information resources to identify and control continuously developing cyber risks. 
The MITRE Corporation, for instance, has created a collection of resources, such as Tactics Techniques and Common Knowledge
repository (ATT\&CK) ~\cite{ATTACK}, Common Attack Pattern Enumeration and Classification (CAPEC)~\cite{CAPEC}, Common Weakness Enumeration (CWE)~\cite{CWE}, and Common Vulnerabilities and Exposures (CVE)~\cite{CVEdataset}. ATT\&CK contains descriptions of attacks as Tactics, Techniques, and Procedures (TTP). CAPEC is called an Attack Pattern, which is ``the common approach and attributes related to exploiting a weakness in a software, firmware, hardware, or service component''~\cite{CAPEC}. Additionally, CAPEC is structured as a list of information about the attack, including techniques for the attack, potential outcomes, and mitigations~\cite{refat2024comparison}.
CWE is a community-developed collection of typical weaknesses in software, coding errors, and security flaws. 
% In cyberattacks, attackers utilize TTPs to compromise the availability, confidentiality, and integrity of the targeted systems. In addition, the MITRE ATT\&CK framework helps to predict and mitigate the TTP that an attacker will employ after the breach. There are a total of 14 tactics, which stand for the \textit{"why"} of behavior. Understanding tactics helps uncover the underlying intentions or objectives that drive behavior. Moreover, 625 techniques represent the \textit{"how"}:  the practical means through which individuals implement their tactics. 809 procedures represent the step-by-step instructions or processes involved in executing behavior.
% CAPEC is called an Attack Pattern, which is ``the common approach and attributes related to exploiting a weakness in a software, firmware, hardware, or service component'' ~\cite{CAPEC}. Additionally, CAPEC is structured as a list of information about the attack, including techniques for the attack, potential outcomes, and mitigations.
CVE is a standardized dictionary of common terms for publicly known cybersecurity vulnerabilities. Its initiative aims to enhance the efficiency of identifying, finding, and fixing software vulnerabilities by providing a unified naming system~\cite{WhatCVE}~\cite{othman2024vulnerability}. The CVE entries are widely used, but CVE issues often lack information regarding mitigation techniques or existing defense strategies that could effectively address a specific vulnerability~\cite{sonmez2021classifying,elder2022really}.
% software weakness gives an attacker unrestricted access to a system or network. Hence, the attacker might be able to assume the identity of a superuser or system administrator and get complete access ~\cite{elder2022really}.
Linking ATT\&CK to CVE issues will help the cyber community, as these two repositories are currently separated. However, linking 625 Attack Techniques manually with 295,604 CVE issues ~\cite{CVEdataset} is a non-trivial task requiring automated models to link attack information with vulnerabilities and weaknesses. 
% In addition, it would be advantageous and time-efficient to begin pre-labeling with machine-learning models before expert validation.
This study aims to develop an approach that links the Attack Techniques (AT) in the ATT\&CK repository to reports in the CVE repository. To achieve this goal, we developed a tool named Automated Vulnerability Detection From Attack Technique Text (VULDAT), that leverages the sentence transformer (MPNET)~\cite{multi-qa-mpnet-base-dot-v1} that connects the textual description of an AT to the textual description of CVE reports. 
% Moreover, to check the effectiveness of VULDAT, we have built an annotated dataset with explicit links found in MITRE repositories. Our approach's dataset and source code are available on Github~\cite{VULDAT}. 
Thus, we aim to answer the following questions:
\enlargethispage{\baselineskip}
\par\noindent
\begin{itemize}
    \item \textit{\textbf{RQ1: } What is the performance of sentence transformer models to detect software vulnerabilities from attack
text?}  
% To provide a solution to this question, we examined the machine learner's performance when evaluated using different attack technique information types. The aim is to find the description of attack techniques of an attack that is more closely related to a vulnerability. 
 \item \textit{\textbf{RQ2: } How many CVE issues VULDAT correctly detect from an attack text?}
 
 % To solve this question, we evaluated the overlap between actual links from our mapping and our detection list of CVE issues generated by VULDAT. The aim is to show VULDAT's effectiveness in detecting the CVE issues related to the textual description of attack text and propose some of the missing links in the MITRE board. 
\end{itemize}
The contributions of this work are as follows: 
% Firstly,
% the proof-of-concept VULDAT supports cybersecurity by providing an automated way of detecting software vulnerabilities from attack text.
% Additionally,
% a novel annotated mapping dataset ~\cite{VULDATDataSet} explicitly links ATT\&CK with vulnerabilities found in MITRE repositories.
% Moreover, an analysis of sentence transformer models highlights their comparative features and performance metrics.
\par\noindent
\begin{itemize}

    \item The proof-of-concept VULDAT supports cybersecurity by providing an automated way of detecting software vulnerabilities from AT text;
    % \item A tool utilizing Transformer-based models and traditional machine learning to link ATT\&CK to CVE issues;
    \item A novel annotated mapping dataset ~\cite{VULDATDataSet} explicitly linking ATT\&CK with vulnerabilities found in MITRE repositories;
    \item  An analysis of sentence transformer models, highlighting their comparative features and performance metrics;
    % \item A list  of missing links between known ATT\&CK and vulnerabilities;

\end{itemize}

% The contributions of this work are as follows. First, linking ATT\&CK and CVE, the experiments utilizing Transformer-based models, traditional machine learning, and data augmentation techniques to link ATT\&CK to CVEs. Second, identifying which attack information is most effective, thoroughly evaluating the most effective model for utilizing the appropriate type of attack information to link ATT\&CK and CVE.

The structure of this paper is as follows. Section \ref{Sec:METHODOLOGY} illustrates our proposed methodology for linking attacks to CVE. Section \ref{Sec:results} summarizes the results of our work. Section \ref{Sec:limitation} outlines the limitations of our work. Finally, section \ref{Sec:conclusion} presents our conclusions and further research.

\begin{figure}[hbt]
    \centering    \includegraphics[width=1.08\linewidth]{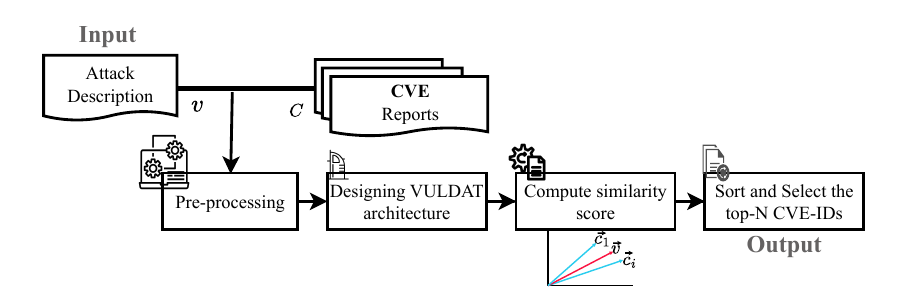}
    \caption{Methodology Overview}\label{fig:OverviewMethidology}
\end{figure}

\section{Study Design and Methodology }
\label{Sec:METHODOLOGY}

In this section, we describe our research methodology for vulnerability detection. Fig.~\ref{fig:OverviewMethidology} illustrates the implemented pipeline.

% \subsection{CVE issues extraction pipeline}
% \label{subsec:ttpextraction}
% We build a CVE extraction process using the four procedures listed below. 

% \textbf{Step 1:} collection of datasets, we collect the dataset (Section \ref{sec:vuldataset}) for the experiment. The dataset includes three types of textual descriptions: (a) textual descriptions of CVE issues, (b) textual descriptions of attack text, and (c) ground truth for the detection task, which are the relevant attacks employed in the CVE.

% \textbf{Step 2}: text processing, we apply the following on the corpus in this stage: Pre-processing includes the following steps: (a) tokenization; (c) removing HTTP links; and (d) removing noise (Notes). 

% \textbf{Step 3 and Step 4:} sentence transformer leverages pre-trained language models such as natural language inference or semantic textual similarity. Hence, in this step, to extract textual features, we use the sentence transformer techniques that we would provide to calculate the similarity score. In the four subsections that follow, we instantiate these four phases. Fig.~\ref{fig:OverviewMethidology} illustrates the implemented pipeline.

%%\vspace{-1cm}
\subsection{Data Collection}
 \label{sec:vuldataset}
 \enlargethispage{\baselineskip}
Table~\ref{tab:golddataset} and Fig.~\ref{fig:MitreConnection} provide an overview of the Mapping dataset $\mathcal{M}$ used in this paper and summarize the number of collected items from MITRE repositories, ATT\&CK ~\cite{ATTACK}, CAPEC ~\cite{CAPEC}, CWE ~\cite{CWE}, and CVE ~\cite{CVEdataset}. ATT\&CK and CAPEC reports are connected through the technique ID, CAPEC and CWE reports are connected through the CWE-ID, and CWE and CVE reports are connected through the CVE-ID.
The dataset provides extensive vulnerability information for effective cybersecurity vulnerability management and prioritization.
% The dataset contains extensive vulnerability information and various databases related to cybersecurity vulnerabilities. Moreover, it is designed to provide a comprehensive source of information to aid in vulnerability management and prioritization for users in the field of cybersecurity.
%\vspace{-0.45cm}
\enlargethispage{\baselineskip}
\begingroup
\setlength{\abovedisplayskip}{0pt}
\setlength{\belowdisplayskip}{0pt}
 \begin{table}[htb]
\caption{ATT\&CK linked and not linked to vulnerability reports\label{tab:golddataset}}
\scriptsize % Smaller font size
\setlength{\tabcolsep}{4pt} % Reduce column padding
\centering
% \bgroup
% \def\arraystretch{1}% 1 is the default, change whatever you need
% \setlength{\tabcolsep}{9pt}
\begin{tabular}{lccc}
\hline  
&\textbf{Linked} &\textbf{Not linked} & \textbf{Total} \\
\hline
\textbf{Attack Patterns} &177 &382 &559\\
\textbf{CWE reports} &117 &817 &934\\
\textbf{CVE reports} &685 & 294919&295604\\
\hline
%\vspace{-0.6cm}
\end{tabular}%
\end{table}
\endgroup\unskip
For instance, we found that only 100 attack techniques/sub-techniques are linked to 88 CWE reports and the same CWE reports are connected to 685 CVE reports. Meanwhile, a substantial number of 293,589 CVE issues, 370 CWEs, 282 CAPEC, and 230 Techniques are isolated and are not linked to any other information, as shown in Fig.~\ref{fig:MitreConnection}.
% shows how many ATT\&CK, CAPEC, CWEs, and CVEs are linked and not linked, which are 382, 356, and 293,589, respectively.
%\vspace{-0.5cm}
\begingroup
\setlength{\abovedisplayskip}{0pt}
\setlength{\belowdisplayskip}{0pt}
\begin{figure}[htb]
\centering
\includegraphics[width=1.05\columnwidth]{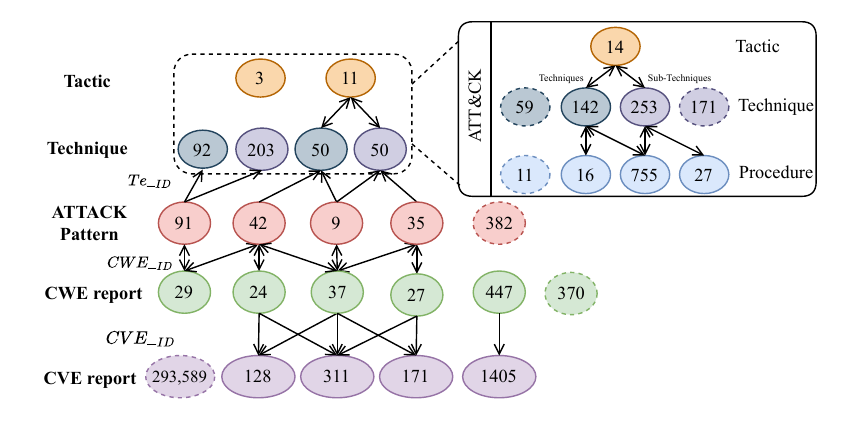}
% \caption*{VULDAT toolbox.}\label{fig:pipeline}
\caption{MITRE repositories and their connections}\label{fig:MitreConnection}
\end{figure}
\endgroup\unskip

%\vspace{-0.5cm}
\enlargethispage{\baselineskip}\textbf{}
\subsection{Pre-Processing}
\label{Sec:preprocessing}
% After the data collection process, we start by removing the citations and URLs. Then, we tokenize the sentence using the \texttt{nltk.stem} from the \texttt{nltk} Python package.
Fig.~\ref{fig:textpre-processing} illustrates the steps of pre-processing.
The pre-processing phase in our experiment involves two parallel steps: (1) 
Our initial pre-processing (partial pre-processing) experiment includes the same steps as the pretrained model: tokenization, stemming, lemmatization, stop word removal, and punctuation handling.
% the original pre-processing of the trained model (partial pre-processing) in our experiment incorporates the same pre-processing steps as those automatically executed by the pretrained model tokenization, stemming, lemmatization, stop word removal, and handling punctuation. 
(2) Additional pre-processing that removes citations, URLs, extra spaces, and non-alphanumeric characters (full pre-processing). Tokenization involves splitting the text into tokens or single words. This process allows us to handle the text more effectively, making extracting meaningful insights and patterns from the data easier. 
Stemming and Lemmatization are the process of breaking down a word into its stem or root form. This process aims to unify variations of words that share the same core meaning, reducing the complexity of the data and improving our analysis.
Moreover, stop words are removed from the textual descriptions of the attack information, as they are often considered irrelevant or noisy for natural language processing tasks. 
We remove citations, URLs, extra spaces, and non-alphanumeric characters from AT texts to guarantee a cleaner dataset.
% However, we found that this pre-processing step adversely affected the quality of the results of our VULDAT model for vulnerability detection. Therefore, we decided to keep the stop words in our dataset.
%\vspace{-0.5cm}
\begingroup
\setlength{\abovedisplayskip}{0pt}
\setlength{\belowdisplayskip}{0pt}
\begin{figure}[htb]
\centering
\includegraphics[width=0.8\columnwidth]{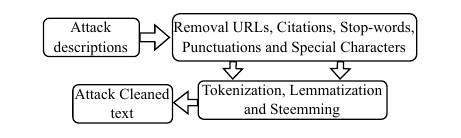}
% \captionsetup{font=small} % Set the font 
\caption{Text pre-processing\label{fig:textpre-processing}}
% %\vspace{-0.3cm}
\end{figure}
\endgroup
\unskip
%\vspace{-0.5cm}
\enlargethispage{\baselineskip}
\subsection{Designing VULDAT Architecture}
\label{sec:similarity}

% In this paper, we use the pre-trained sentence transformer models ~\cite{pretrained}, based on the Distilbert, Bert, Roberta, and Mpnet architectures. In addition, they are trained on multilingual data and can compute sentence embeddings for different languages and cross-lingual tasks. These models can convert natural language sentences and paragraphs into numerical vectors, capturing the meaning and similarity of the texts. Moreover, these vectors can be useful for tasks like grouping similar texts and searching for related documents. The model is a new version of the BERT model ~\cite{reimers2019sentence} for natural language comprehension. It employs a 12-layer neural network to generate vectors that encapsulate the text. Unlike BERT, which learns from a single language, this model is trained on a diverse data set from 50 different languages.
In this paper, we use state-of-the-art pre-trained sentence transformer models~\cite{pretrained} as summarized in Table~\ref{tab:pretrainedmodels}. These models are based on the Distilbert ~\cite{sanh2019distilbert}, Bert~\cite{devlin2018bert}, Roberta~\cite{liu2019roberta}, and Mpnet~\cite{song2020mpnet} architectures. These include \texttt{multi-qa-MiniLM-L6-cos-v1} and \texttt{all-MiniLM-L12-v2}, which utilize the MiniLM architecture for multilingual and cross-lingual tasks.
%\vspace{-0.4cm}
\begingroup
\setlength{\abovedisplayskip}{0pt}
\setlength{\belowdisplayskip}{0pt}
\begin{table}[htb]
\caption{Pretrained Model Summary}
\label{tab:pretrainedmodels}
\scriptsize % Smaller font size
\setlength{\tabcolsep}{2pt} % Reduce column padding
\centering
\bgroup
\begin{tabular}{lccc}
\hline  
\textbf{Model} & \textbf{Embedding Dimension}  & \textbf{Model Size} \\
\hline
multi-qa-MiniLM-L6-cos-v1  & 384  & 80 \, MB \\
multi-qa-distilbert-cos-v1  & 768 & 250 MB \\
all-MiniLM-L12-v2  & 384 & 120 MB \\
all-distilroberta-v1  & 768  & 290 MB \\
multi-qa-mpnet-base-dot-v1  & 768  & 420 MB \\
all-MiniLM-L6-v2  & 384 & 80 \, MB \\
paraphrase-multilingual-MiniLM-L12-v2  & 384  & 420 MB \\
all-mpnet-base-v2  & 768 & 420 MB \\
paraphrase-MiniLM-L6-v2   & 384& 61 \, MB \\
\hline
\end{tabular}%
\egroup
\end{table}
\endgroup\unskip
\enlargethispage{\baselineskip}
% \begingroup
% \setlength{\abovedisplayskip}{0pt}
% \setlength{\belowdisplayskip}{0pt}
% \begin{table}[htb]
% \caption{Pretrained Model Summary}
% \label{tab:pretrainedmodels}
% \centering
% \bgroup
% \def\arraystretch{1}% 1 is the default, change whatever you need
% \setlength{\tabcolsep}{2pt} % Adjust this value as needed
% \begin{tabular}{lcc}
% \hline  
% \textbf{Model} & \textbf{Architecture} & \textbf{Embedding Dimension} \\
% \hline
% MQ MiniLM L6 v1 & MiniLM & 384 \\
% MQ DistilBERT v1 & DistilBERT & 768 \\
% All MiniLM L12 v2 & MiniLM & 384 \\
% All DistilRoBERTa v1 & DistilRoBERTa & 768 \\
% MQ MPNet Base v1 & MPNet & 768 \\
% All MiniLM L6 v2 & MiniLM & 384 \\
% Paraphrase ML MiniLM L12 v2 & MiniLM & 384 \\
% All MPNet Base v2 & MPNet & 768 \\
% Paraphrase MiniLM L6 v2 & MiniLM & 384 \\
% \hline
% \end{tabular}%
% \egroup
% \end{table}
% \endgroup\unskip
%\vspace{-.2cm}
Furthermore, \texttt{multi-qa-distilbert-cos-v1} and \texttt{all-distilroberta-v1}, which are based on the DistilBERT and DistilRoBERTa architectures, respectively. Using the MPNet architecture, the \texttt{multi-qa-mpnet-base-dot-v1} and \texttt{all-mpnet-base-v2}. Lastly, the MiniLM architecture, \texttt{all-MiniLM-L6-v2}, \texttt{paraphrase-multilingual-MiniLM-L12-v2}, and \texttt{paraphrase-MiniLM-L6-v2} models, are designed for use in multilingual paraphrasing tasks. 
% These models are a new version of the BERT model~\cite{reimers2019sentence} for natural language comprehension. Moreover, they employ a 6 and 12-layer neural network to generate vectors that encapsulate the text. Unlike BERT, which learns from a single language, this model is trained on a diverse data set from 50 different languages.
% XXX add a layer for computing cosinus similarity between tech text and CVE reports. Cosine similarity score measures how similar the text descriptions of CVE issues and attack text are and assign the most relevant CVEs to attack.
The sentence transformer models generate embeddings for the AT and the CVE descriptions. Then, we calculate a cosine similarity score between these vectors, which measures how closely the CVE descriptions match the AT text. Thus, the similarity score helps identify and assign the most relevant CVEs to each AT.
% %\vspace{-0.2cm}
\subsection{Compute Similarity Score}
% This can assist us in comprehending the vulnerability's possible effects, how to mitigate their impact, and the threat actors and malicious activities that use them. 
% We apply our model to associate an AT to  CVE reports through their similarity score. The output is ranked among the CVE reports based on their similarity with an AT vector. The top N CVE reports that have similarity greater than 60\% are then associated with the AT vector. The set of these CVE reports is called the Detection list ($\mathcal{L}$). 
We apply our model to associate an AT with CVE report vectors by leveraging their cosine similarity scores. The output CVE vectors are ranked based on their similarity to the AT vector. 
The top N CVE reports with a similarity score greater than 60\% are then associated with the AT vector. This set of CVE reports, named the Detection List ($\mathcal{L}$), represents the most relevant vulnerabilities linked to the AT.

% This threshold ensures the system identifies the most relevant CVEs related to the attacks. 
% %\vspace{-0.2cm}
% %\vspace{-0.1cm}
% %\vspace{-0.2cm}

%\vspace{-0.2cm}
\subsection{Performance Evaluation}
\label{sec:metrics}
% \enlargethispage{\baselineskip}
To measure the accuracy of VULDAT, we use Precision, Recall, and their harmonic
mean F1 to evaluate its performance. Precision is defined by: $ \textit{\textbf{Precision}} = \frac{\textit{TP}}{\textit{TP} + \textit{FP}}$,
Recall is defined by: $
% \begin{align*}
\textit{\textbf{Recall}} = \frac{\textit{TP}}{\textit{TP} + \textit{FN}}$, and $\textit{\textbf{F1}} = 2 \times \frac{\textit{Precision} \times \textit{Recall}}{\textit{Precision} + \textit{Recall}}$. Table~\ref{tab:metricsPreformance2} summarizes the detected positives, negatives, and the different classification's output.
%\vspace{-0.5cm}
\begingroup
\setlength{\abovedisplayskip}{0pt}
\setlength{\belowdisplayskip}{0pt}
\begin{table}[htb]
\scriptsize
\centering
\caption{Positives and Negatives. $\mathcal{A}$ = Set of all Attacks, C = Set of all CVE issues}
\label{tab:metricsPreformance2}
\begin{tabular}{@{}ll@{}}
\hline
\textbf{Type}  & \textbf{Description}   \\ 
\hline
Positives & $\{ a \in \mathcal{A} \, : \exists \, {c} \in \mathcal{M}_{a} \}$\\

Negatives &  $\{ a \in \mathcal{A} \, : \nexists \, {c} \in \mathcal{M}_{a} \}$\\
True Positive  & $\{ a \in \mathcal{A} \, : \left| \mathcal{L}_{a} \cap \mathcal{M}_{a} \right| > 0  \} $\\
False Positive   & $\{ a \in \mathcal{A} \, :\left( \left| \mathcal{L}_{a} \right| > 0 \,\land\, \left|\mathcal{M}_{a} \right| = 0 \right)\}$ \\
False Negative   & $\{ a \in \mathcal{A} \, : \left( \left| \mathcal{L}_{a} \right| = 0 \,\land\, \left| \mathcal{M}_{a} \right| > 0 \right)\}$ \\
True Negative   & $\{a \in \mathcal{A} \, : \left( \left| \mathcal{L}_{a} \right| = 0 \,\land\,  \left| \mathcal{M}_{a} \right| = 0 \right)\}$ \\
\hline
\end{tabular}
\end{table}
\endgroup
\unskip
% %\vspace{-0.2cm} 
\textit{Positives} is when there is an explicit link between an AT text and a CVE, which means when we have $\mathcal{L}$ is not empty, while a \textit{negatives} is when $\mathcal{L}$ is empty which means no explicit link exists.

\enlargethispage{\baselineskip}
% %\vspace{-0.2cm}
\section{Results}
\label{Sec:results}
This section presents the outcomes for each of the RQs.
\par\noindent
\textbf{RQ1}: What is the performance of sentence transformer models to detect software vulnerabilities from attack
text?

% To answer RQ1, we evaluated the performance of the sentence transformer models in detecting CVE reports based on the information from the ATT\&CK repository. 
% Table ~\ref{tab:metricsPerformance} illustrates the precision, recall, and F1. Precision measures the ratio of AT vectors that  at least one CVE, while recall measures the ratio of all attack texts having at least one CVE that are correctly identified by VULDAT. The F1 score is the harmonic mean of precision and recall.

To answer RQ1, we evaluated the performance of the sentence transformer models in detecting CVE reports based on the techniques information from the ATT\&CK repository under different pre-processing conditions (see Section \ref{Sec:preprocessing}). 
Table ~\ref{tab:metricsPerformance} illustrates the precision, recall, and F1. 
% Precision measures the ratio of attack text that has at least one CVE, while recall measures the ratio of all attack texts having at least one CVE that are correctly identified by VULDAT. The F1 score is the harmonic mean of precision and recall. 
Among the models evaluated, the \texttt{multi-qa-mpnet-base-dot-v1} model achieved the best performance metrics under both partial and full pre-processing. With partial pre-processing, it demonstrated a precision of 0.73, a recall of 0.87, and an F1 score of 0.79. Furthermore, under full pre-processing, the model demonstrated significant improvements across all metrics,  with an F1 score of 0.85, precision of 0.86, and recall of 0.83.
However, the \texttt{multi-qa-distilbert-cos-v1} model performed relatively poorly in comparison
under either partial pre-processing with an F1 score of 0.32 or full 
% \enlargethispage{\baselineskip}
%\vspace{-0.5cm}
\begin{table}[htb]
\scriptsize % Smaller font size
\setlength{\tabcolsep}{1.5pt} % Reduce column padding
\centering
\caption{Performance metrics of sentence transformer Models}
\label{tab:metricsPerformance}
\begin{tabular}{lccc|ccc}
\hline
 & \multicolumn{3}{c}{\textbf{Partial pre-processing}} & \multicolumn{3}{c}{\textbf{Full pre-processing}} \\
\cline{2-4} \cline{5-7}
\textbf{Model} & \textbf{Precision} & \textbf{Recall} & \textbf{F1} & \textbf{Precision} & \textbf{Recall} & \textbf{F1} \\ 
\hline
multi-qa-MiniLM-L6-cos-v1 & 1.00 & 0.32 & 0.48 & 1.00 & 0.26 & 0.42 \\
multi-qa-distilbert-cos-v1& 1.00 & 0.19 & 0.32 & 0.88 & 0.13 & 0.22 \\
all-MiniLM-L12-v2 & 0.87 & 0.62 & 0.72 & 0.91 & 0.54 & 0.68 \\
all-distilroberta-v1 & 0.63 & 0.60 & 0.62 & 0.69 & 0.74 & 0.71 \\
\textbf{multi-qa-mpnet-base-dot-v1} & \textbf{0.73} & \textbf{0.87 }& \textbf{0.79} & \textbf{0.86} & \textbf{0.83} & \textbf{0.85} \\
all-MiniLM-L6-v2 & 0.95 & 0.38 & 0.54 & 0.74 & 0.26 & 0.38 \\
paraphrase-multilingual-MiniLM-L12-v2 & 0.67 & 0.85 & 0.75 & 0.65 & 0.93 & 0.76 \\
all-mpnet-base-v2 & 0.75 & 0.77 & 0.76 & 0.81 & 0.80 & 0.81 \\
paraphrase-MiniLM-L6-v2 & 0.68 & 0.46 & 0.55 & 0.80 & 0.61 & 0.69 \\
\hline
\end{tabular}
\end{table}
\vspace{-0.1cm}
\enlargethispage{\baselineskip}
pre-processing with an F1 score of 0.22. In our case, the size of the model is associated with the number of its parameters and may have an impact on the performance of the models, as shown in Table \ref{tab:pretrainedmodels}. A larger size allows the model to capture more complex patterns.
For example, the \texttt{multi-qa-mpnet-base-dot-v1} model has more parameters with a size of 420 MB, which allows it to work more efficiently.
On the other hand, smaller models like the \texttt{multi-qa-distilbert-cos-v1}, which is 250 MB in size, have fewer parameters, which could lower performance. Thus, the F1 scores are 0.79 and 0.32 for partial pre-processing and 0.85 and 0.22 for full pre-processing.
\par\noindent
\textbf{RQ2}: How many CVE issues VULDAT correctly detect from an attack text?
\enlargethispage{\baselineskip}
% %\vspace{-0.1cm}
To answer RQ2, we compute the Jaccard Similarity, Mapping Accuracy, and Detection Accuracy for each AT text. 
The Jaccard Similarity evaluates the similarity of two sets: the detected set ($\mathcal{L}_{a}$) by VULDAT  and actual sets  ($\mathcal{M}_{a}$).
It measures the ratio of the intersection of these sets to their union,
$
\mathrm{Jaccard \, Similarity }= \frac{{|\mathcal{L}_{a} \cap \mathcal{M}_{a}|}}{|\mathcal{L}_{a} \cup \mathcal{M}_{a}|}
$.
Mapping Accuracy is the ratio of the intersection of the detected set ($\mathcal{L}_{a}$) and the actual set ($\mathcal{M}_{a}$) to the size of the actual set ($\mathcal{M}_{a}$), 
$
\mathrm{Mapping \, Accuracy} = \frac{{|\mathcal{L}_{a} \cap \mathcal{M}_{a}|}}{{|\mathcal{M}_{a}|}}
$.
% \enlargethispage{\baselineskip}
Detection Accuracy is the ratio of the intersection of the detected set ($\mathcal{L}_{a}$) and the actual set ($\mathcal{M}_{a}$) to the size of the detected set ($\mathcal{L}_{a}$), $
\mathrm{Detection \, Accuracy} = \frac{{|\mathcal{L}_{a} \cap \mathcal{M}_{a}|}}{{|\mathcal{L}_{a}|}}$. A higher Detection Accuracy implies more accurate identifications by VULDAT.
% Fig.~\ref{fig:alljaccardplot} illustrates the Jaccard Similarity among sentence transformer models for all attack texts, indicating that the \texttt{multi-qa-mpnet-base-dot-v1} model achieves the highest Jaccard Similarity.  Thus, we selected the multi-qa-mpnet-base-dot-v model to compute the Jaccard Similarity, Mapping Accuracy, and Detection Accuracy. 
%\vspace{-.3cm}
\begin{figure}[htb!]
\centering
\includegraphics[width=.7\columnwidth]{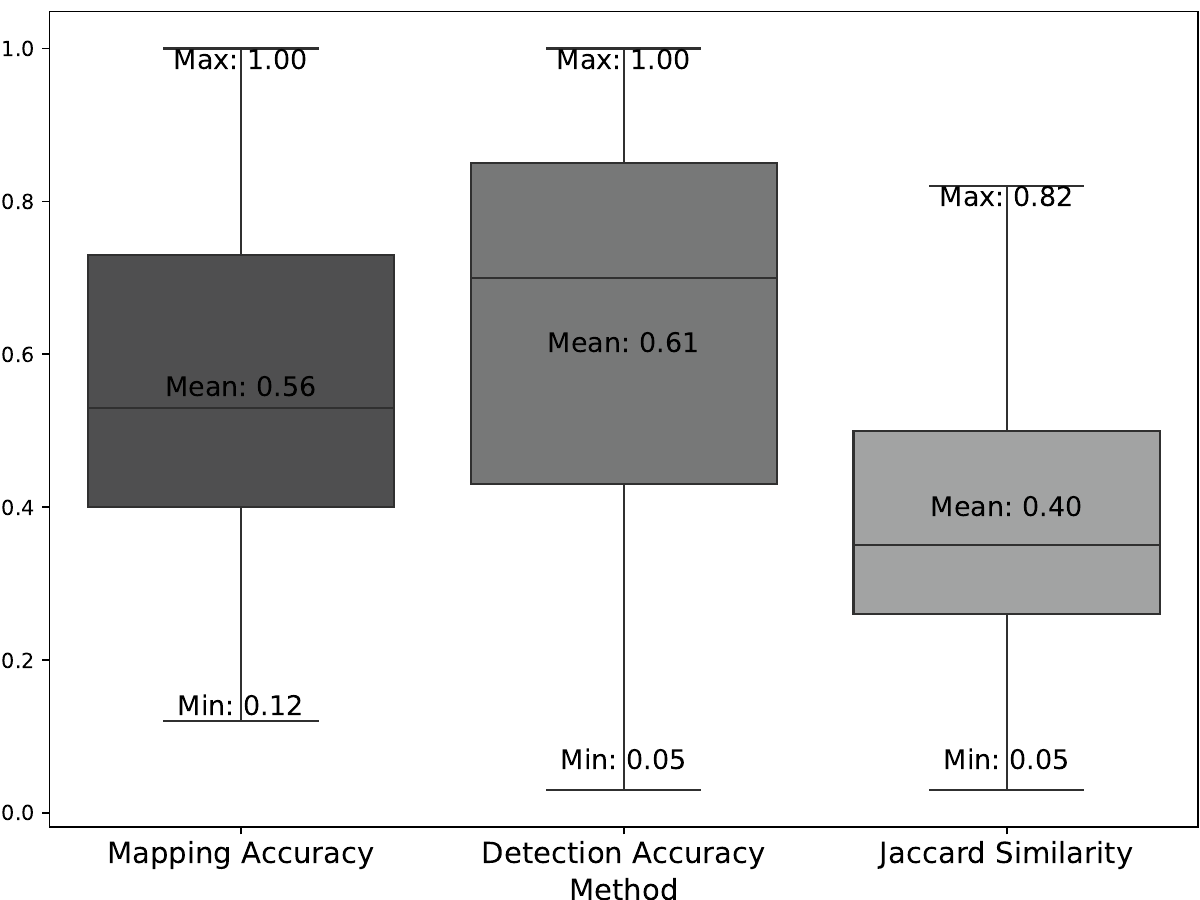}
% \captionsetup{font=small} % Set the font size of the caption to small
\caption{VULDAT performance}\label{fig:jaccardplot}
%\vspace{-.3cm}
\end{figure}
% \enlargethispage{\baselineskip}
Fig.~\ref{fig:jaccardplot} reports the distribution of these measures across all AT texts, showing a Mapping Accuracy of 0.56 and VULDAT's reported Detection Accuracy of 0.61. Moreover, 
VULDAT reports a Jaccard Similarity of 0.40 on average, with a range spanning from 0.05 to 0.82.
Whiskers extend further below Q1 and above Q3, indicating a slight skewness in the data distribution. 
Notably, most of the similarity scores are higher than 0.50, indicating that many sample comparisons demonstrate relatively high similarity. For instance, the T1539 attack achieves the highest Jaccard Similarity score of 0.82. Specifically, it consists of 150 CVE issues in $\mathcal{L}$, 125 CVE issues in $\mathcal{M}$, with 124 CVE issues in $\mathcal{L}_{a} \cap \mathcal{M}_{a}$, and  151 CVE issues at $\mathcal{L}_{a} \cup \mathcal{M}_{a}$.
\enlargethispage{\baselineskip}
Additionally, a lower Jaccard Similarity value suggests limited agreement between the elements of the real and detected sets. Specifically, T1003 exhibits a Jaccard Similarity of 0.05. Upon manual evaluation of this attack by examining which CVE related to the AT text, VULDAT discovered the size of   $\mathcal{L}_{T1003}$ is 37, and we found 22 of these links do not exist in the MITRE repositories.
% VULDAT discovered the size of   $\mathcal{L}_{T1003}$ is 37 and the size of $\mathcal{L}_{T1003} \cap \mathcal{M}_{T1003}$ is 2, 
% whereas it should have been 24 based on our manual validation. Following this validation process, we identified 22 new CVE links provided by VULDAT that should be included in $\mathcal{M}_{T1003}$.
% %\vspace{-0.4cm}
% %\vspace{-0.2cm}
Thus, it is important to note that many detected
CVE links may indeed be linked to the AT. However, their link is missing from the MITRE repositories. Our future plan to validate all of these $\mathcal{L}_{a}-\mathcal{M}_{a}$ links are indeed connected.

\section{Related Work}
\enlargethispage{\baselineskip}
The Cve2att\&ck ~\cite{grigorescu2022cve2att} model uses BERT-based~\cite{devlin2018bert} language models and classic models for machine learning
to automatically link CVE reports with techniques that are able to link CVE issues with the MITRE ATT\&CK Enterprise Matrix techniques by utilizing the textual description present in CVE metadata. 
% In addition, the study used a three-step CVE extract process to extract the CVE issues published between 2020 and 2022. The first step is to identify the type of CVE (such as buffer overflow or SQL injection). Next, the attacker is able to access which functionality by taking advantage of the CVE. The last step is to identify the exploitation technique, using the tips supplied to provide information about the actions required to exploit a vulnerability.
The CVE Transformer (CVET) ~\cite{ampel2021linking} model uses RoBERTa~\cite{liu2019roberta} to link a CVE to ten tactics from ATT\&CK Enterprise Matrix. Kuppa et al.~\cite{kuppa2021linking} introduced the Multi-Head Joint Embedding Neural Network model to automatically map CVE issues to ATT\&CK techniques. The suggested representation finds particular regular expressions in the current threat reports and then compares the ATT\&CK technique vectors with the text description found in the CVE metadata by measuring their similarity using the cosine distance.
SMET ~\cite{abdeen2023smet} is an automatic mapping tool that matches CVE entries to ATT\&CK techniques using textual similarity. It uses the ATT\&CK BERT model, trained on the SIAMESE network, to determine the semantic similarity between attack actions.
% In this study, the researchers constructed the dataset manually, which had CVE entries published from 2014 to 2022.

% BRON~\cite{hemberg2021linking}  is a bi-directional aggregated data graph that supports path tracing between ATT\&CK Enterprise Matrix tactics and techniques, CWE, CVE, and Attack Patterns. BRON creates a graph framework that connects all of the MITRE information. A link exists between the CVE list and ATT\&CK by following the relational linkages within the created graph. However,  the model fails as it cannot link the most recent CVE reports to ATT\&CK Enterprise Matrix techniques. Additional works that used of a CVE dataset from BRON~\cite{hemberg2021linking} include  ~\cite{liu2023not}, ~\cite{ampel2024improving},  ~\cite{hemberg2021using}, and ~\cite{ampel2021linking}. This dataset offers classifications into higher-level abstractions, such as MITRE ATT\&CK techniques and tactics. For threat intelligence and modeling, other works, such as  ~\cite{fairbanks2021att}, ~\cite{fairbanks2021identifying}, and ~\cite{dravsar2020session}, used ATT\&CK and graphs.

%\vspace{-0.2cm}
\section{Threats to Validity}
\label{Sec:limitation}
In this section, we discuss the limitations of our research.
\par\noindent
\textbf{Construct validity.} We determine the output of VULDAT by experimentally setting a 60\% threshold. We plan to perform a sensitivity analysis on the threshold to substantiate our choice. 
%For example, we may have different degrees of similarity since we have different types of attack texts and CVE descriptions. 
%A second threat concerns the size of the dataset. Our dataset contains 100 attack techniques, 685 linked CVE issues  out of 293,000 existing in the MITRE repository.
\par\noindent
\textbf{Internal validity.}
We exploit the description of the attack techniques. Other attack information included in the MITRE repository could eventually provide better performance results.
% We are working to extend our work to other types of attack information, such as tactics and procedures from the ATT\&CK repository and attack patterns from the CAPEC repository.

\par\noindent
\textbf{External validity.}
The link between attacks and vulnerabilities is based on what the MITRE repositories explicitly indicate. These links have not been updated with a stringent schedule. Thus, our dataset may not contain the very last pieces of information about an attack and its related vulnerabilities. 
% Therefore, we plan to extend VULDAT to information captured from other sources, such as specialized magazines and issue trackers of publicly available projects (e.g., GitHub issues). 
%\vspace{-0.1cm}
\section{Conclusion}
\label{Sec:conclusion}
% This paper has introduced VULDAT, a new approach to detect CVE issues from textual descriptions of attack behavior, utilizing machine learning for natural language processing methods. VULDAT leverages the MPNET sentence transformer to automatically identify vulnerabilities from the description of attack techniques.
% The MITRE ATT\&CK and the CVE repositories are the input and output data sources for our model. First, we compared VULDAT's performance with nine SOTA models using the description of attack techniques. Second, we evaluated VULDAT's ability to retrieve all vulnerabilities for an attack technique. To this aim, we introduced three accuracy measures: Mapping, Detection, and Jaccard Similarity. The results indicate that VULDAT achieves a Mapping Accuracy of 0.56, Detection Accuracy of 0.61, and Jaccard Similarity of 0.40. Moreover, VULDAT archives an F1 score of 0.85, a Precision of 0.86, and a Recall of 0.83. 
% By manually inspecting VULDAT's output for one attack technique, we realized that some of the vulnerabilities identified by VULDAT must also have been associated with the attack in the CVE repository.
% Our approach can provide a new contribution to the cybersecurity community by creating an automatic link between attack and vulnerability knowledge bases. 

This paper introduces VULDAT, an approach that uses MPNET to detect CVE issues from textual descriptions of attack behaviors. It utilizes data from MITRE ATT\&CK and CVE repositories. VULDAT’s performance was compared with nine state-of-the-art models and evaluated using three accuracy measures: Mapping, Detection, and Jaccard Similarity. VULDAT achieved a Mapping Accuracy of 0.56, a Detection Accuracy of 0.61, a Jaccard Similarity of 0.40, an F1 score of 0.85, a Precision of 0.86, and a Recall of 0.83. Manual inspection of its output confirmed some identified vulnerabilities were associated with attacks in the CVE repository.
We plan to investigate other models (e.g., sequence-to-sequence models) for detecting CVE issues from attack text. Furthermore, we will perform a large manual inspection of VULDAT output to recommend missing links between attacks and vulnerabilities to the MITRE board. 
% %\vspace{-0.3cm}
\enlargethispage{\baselineskip}
\section{Acknowledgements}
The first author thanks the CSLab at the Free University of Bozen-Bolzano for supporting this work under project no. EFRE1039 in the EFRE-FESR 2021-2027 program.
\enlargethispage{\baselineskip}

% %\vspace{-0.5cm}
% \bibliographystyle{unsrt}
\bibliographystyle{IEEEtran}
\bibliography{sample-base}
\end{document}